# Room-temperature ferrimagnetism of anti-site-disordered $Ca_2MnOsO_6$

Hai L. Feng (冯海),[1,2,3,*,#] Madhav Prasad Ghimire,[4,5,6] Zhiwei Hu,[2] Sheng-Chieh Liao,[2] Stefano Agrestini,[2,7] Jie Chen,[1,8] Yahua Yuan,[9,1] Yoshitaka Matsushita,[10] Yoshihiro Tsujimoto,[1,8] Yoshio Katsuya,[11] Masahiko Tanaka,[11] Hong-Ji Lin,[12] Chien-Te Chen,[12] Shih-Chang Weng,[12] Manuel Valvidares,[7] Kai Chen,[13] Francois Baudelet,[13] Arata Tanaka,[14] Martha Greenblatt,[3] Liu Hao Tjeng,[2] Kazunari Yamaura [1,8*]

1. *Research Center for Functional Material, National Institute for Materials Science, 1-1 Namiki, Tsukuba, Ibaraki 305-0044, Japan*
2. *Max Planck Institute for Chemical Physics of Solids, Nöthnitzer Str. 40, 01187 Dresden, Germany*
3. *Department of Chemistry and Chemical Biology, Rutgers, the State University of New Jersey, 610 Taylor Road, Piscataway, New Jersey 08854, United States*
4. *Central Department of Physics, Tribhuvan University, Kirtipur, 44613 Kathmandu, Nepal*
5. *Leibniz Institute for Solid State and Materials Research, IFW Dresden, P.O. Box 270116, D-01171 Dresden, Germany*
6. *Condensed Matter Physics Research Center, Butwal-11, Rupandehi, Nepal*
7. *ALBA Synchrotron Light Source, E-08290 Cerdanyola del Vall`es, Barcelona, Spain*
8. *Graduate School of Chemical Sciences and Engineering, Hokkaido University, North 10 West 8, Kita-ku, Sapporo, Hokkaido 060-0810, Japan*
9. *School of Physics and Electronics, Central South University, Changsha 410083, China*
10. *Materials Analysis Station, National Institute for Materials Science, 1-2-1 Sengen, Tsukuba, Ibaraki 305-0047, Japan*
11. *Synchrotron X-ray Station at SPring-8, National Institute for Materials Science, Kouto 1-1-1, Sayo-cho, Hyogo 679-5148, Japan*
12. *National Synchrotron Radiation Research Center, 101 Hsin-Ann Road, Hsinchu 30076, Taiwan, ROC*
13. *Synchrotron SOLEIL, L'Orme des Merisiers, Saint-Aubin, 91192 Gif-sur-Yvette Cedex, France*
14. *Department of Quantum Matter, ADSM, Hiroshima University, Higashi-Hiroshima 739-8526, Japan*

* Corresponding authors E-mail: hai.feng@iphy.ac.cn, Hai.Feng_nims@hotmail.com (HLF); Yamaura.Kazunari@nims.go.jp (KY)

# Current address: The Institute of Physics, Chinese Academy of Sciences, Beijing 100190, China

**Abstract**

Room-temperature ferrimagnetism was discovered for the anti-site-disordered perovskite $Ca_2MnOsO_6$ with $T_c$ = 305 K. $Ca_2MnOsO_6$ crystallizes into an orthorhombic structure with a space group of *Pnma*, in which Mn and Os share the oxygen-coordinated-octahedral site at an equal ratio without a noticeable ordered arrangement. The material is electrically semiconducting with variable-range-hopping behavior. X-ray absorption spectroscopy confirmed the trivalent state of the Mn and the pentavalent state of the Os. X-ray magnetic circular dichroism spectroscopy reveals that the Mn and Os magnetic moments are aligned antiferromagnetically, thereby classifying the material as a ferrimagnet which is in accordance with band structure calculations. It is intriguing that the magnetic signal of the Os is very weak, and that the observed total magnetic moment is primarily due to the Mn. The $T_c$ = 305 K is the second highest in the material category of so-called disordered ferromagnets such as $CaRu_{1-x}Mn_xO_3$, $SrRu_{1-x}Cr_xO_3$, and $CaIr_{1-x}Mn_xO_3$, and hence, may support the development of spintronic oxides with relaxed requirements concerning the anti-site disorder of the magnetic ions.

# 1. INTRODUCTION

Double perovskite oxides containing 3d and 4d/5d elements are currently much in focus since promising properties for spintronic applications have been reported. For example, $Sr_2FeMoO_6$ shows a low-field magnetoresistance at room temperature[1], and $Sr_2CrReO_6$ half-metallic (HM) transport with a remarkably high Curie temperature ($T_c$) of 635 K[2,3]. Analogous closely related ferrimagnetic (FIM) oxides have been synthesized with a wide variety of 3d and 4d/5d elements, such as $A_2FeMoO_6$ ($A$ = Ca, Ba)[1,4], $A_2FeReO_6$ ($A$ = Ca, Sr, Ba)[5,6], $Ca_2CrOsO_6$[2,3], and $Ca_2FeOsO_6$[7,8]. In addition, a ferromagnetic (FM) Dirac–Mott insulating state ($T_c$ ~100 K) has been found in $Ba_2NiOsO_6$ [9] and an exchange bias effect in $Ba_2Fe_{1.12}Os_{0.88}O_6$ [10], which may also be useful for further development of spintronic oxides.

In the magnetic ground state of $Sr_2FeMoO_6$ and $Sr_2CrReO_6$, the 3d magnetic moments are ordered parallel to each other and antiparallel to those of the 4d/5d[1-3]. Anti-site disorder between the 3d and 4d/5d elements has, however, a significant negative impact on the magnetic properties, because strongly antiferromagnetic (AFM) Fe–O–Fe and Cr–O–Cr bonds are formed which interfere with the long-range magnetic order[11,12]. Indeed, even a small degree of anti-site disorder dramatically decreases $T_c$, and the spin-polarization is also strongly reduced in $Sr_2FeMoO_6$[13,14], accompanied by a linearly decreasing saturation magnetization[12]. Therefore, accurate control of the anti-site disorder has been a significant issue for the fabrication of practical devices since growing anti-site-disorder-free materials is highly challenging[11,12]. Alternatively, anti-site-disorder tolerant materials with promising properties are in demand.

A so-called disordered FM has been reported in a substitutional study of the paramagnetic perovskite $CaRuO_3$, where Ru is partially replaced by a variety of magnetic or nonmagnetic elements such as Sn, Ti, Mn, Fe, Ni, or Rh[15,16]. Among these, $Ca_2MnRuO_6$ is particularly of high interest, because its magnetic moment is relatively large (~1.6 $\mu_B$/f.u.) and its $T_c$ is high (~230 K)[17,18]. In the structure of $Ca_2MnRuO_6$, Mn and Ru are distributed over the perovskite $B$ site at an equal ratio without an ordered arrangement. This material was revealed to be FIM by neutron diffraction owing to a balance between FM ($Mn^{3+}$ to $Mn^{4+}$) and AFM (Ru to Mn) interactions[17-19], and the experimental magnetic state was consistent with that obtained by first-principles calculations[19]. Another example is $SrRu_{0.6}Cr_{0.4}O_3$ which was found to have a high transition temperature of 400 K but a very small saturated moment of 0.15 $\mu_B$/f.u.[20].

In this study, an anti-site-disordered compound, $Ca_2MnOsO_6$ was synthesized for the first time by a high-pressure and high-temperature method at 6 GP and 1500 °C. $Ca_2MnOsO_6$ shows a FIM transition at $T_c$ = 305 K, which is the second highest $T_c$ among the disordered FMs. Here we report the refined crystal structure and bulk magnetic properties of $Ca_2MnOsO_6$. The results suggest that the compound is useful for further development of anti-site-disorder-tolerant spintronic oxides.

## 2. EXPERIMENTAL

Polycrystalline $Ca_2MnOsO_6$ was synthesized via a solid-state reaction from powders of $CaO_2$ (lab-made from $CaCl_2 \cdot 2H_2O$, 99% Wako Pure Chem.), Os (99.95%, Heraeus Materials), and $MnO_2$ (99.997%, Alfa-Aesar). The powders were thoroughly mixed at the stoichiometric ratio, followed by sealing in a Pt capsule. The preparation was conducted in an Ar-filled glove box. The Pt capsule was statically and isotropically compressed in a belt-type high-pressure apparatus (Kobe Steel, Ltd., Japan [21]), and a pressure of 6 GPa was continuously applied while the capsule was heated at 1500 °C for 1 h, followed by quenching to room temperature in less than a minute. The pressure was then gradually released over several hours.

A dense, black polycrystalline pellet was obtained, and several pieces were cut out from it. A selected piece was finely ground for a synchrotron X-ray diffraction (SXRD) study, which was conducted in a large Debye–Scherrer camera in the BL15XU beam line, SPring–8, Japan[22,23]. The SXRD pattern was collected at room temperature and the wavelength was confirmed to be 0.65298 Å by measurement of a standard material, $CeO_2$. The absorption coefficient was measured in the same line. The Rietveld method was used to analyze the SXRD pattern with the RIETAN–VENUS software [24,25].

X-ray absorption spectroscopy (XAS) at the Mn-$L_{2,3}$ and Os-$L_3$ edges of $Ca_2MnOsO_6$ was carried out at the BL11A and BL07C beamlines using the total electron yield and transmission method, respectively, in the National Synchrotron Radiation Research Center, Taiwan. The Mn-$L_{2,3}$ spectrum of MnO and the Os-$L_3$ spectrum of $Sr_2FeOsO_6$ were also measured for energy calibration purposes. X-ray magnetic circular dichroism (XMCD) spectra at the Mn-$L_{2,3}$ and the Os-$L_{2,3}$ edges were obtained at the BL29 BOREAS beamline of the ALBA synchrotron radiation facility in Barcelona and at the ODE beamline of Soleil France, respectively. The degree of circular polarization in BOREAS and ODE beamline was close to 100% and 90%, respectively. The XMCD spectra were measured in a magnetic field of 60 kOe at a temperature of 20 K for Mn-$L_{2,3}$ and in 13 kOe at 4 K for the Os-$L_{2,3}$. The Os $L_3/L_2$ edge-jump intensity ratio $I(L_3)/I(L_2)$ was normalized to 2.08 [26], which accounts for the difference in the radial matrix elements of the $2p_{1/2}$–to–$5d(L_2)$ and $2p_{3/2}$–to–$5d(L_3)$ transitions.

The electrical resistivity ($\rho$) of polycrystalline $Ca_2MnOsO_6$ was measured by a four-point method at a gauge current of 0.1 mA in a physical properties measurement system (Quantum Design, Inc.). Electrical contacts on a piece of $Ca_2MnOsO_6$ were prepared using Pt wires and Ag paste in the longitudinal direction. The temperature dependence of the specific heat capacity ($C_p$) was measured in the same apparatus by a thermal relaxation method at temperatures between 2 K and 300 K using Apiezon N grease to thermally connect the material to the holder stage.

The magnetic susceptibility ($\chi$) of a loosely gathered $Ca_2MnOsO_6$ powder was measured in a magnetic properties measurement system (Quantum Design, Inc.). The measurement was conducted in field cooling (FC) and zero-field cooling (ZFC) conditions in a temperature range between 2 K and 390 K. The applied magnetic field was 10 kOe. The magnetic field dependence of the magnetization ($M$) was measured between −50 kOe and +50 kOe at fixed temperatures of 2, 200, 300, and 350 K. The alternative current (ac) $\chi$ was measured in the same apparatus between 5 K and 320 K; the amplitude and frequencies of the ac-magnetic field were 5 Oe and 0.5–500 Hz, respectively.

The density functional theory (DFT) calculation was performed on $Ca_2MnOsO_6$ with the all-electron full-potential local-orbital code[27] using the standard generalized-gradient approximation (GGA) [28]. In this study, a linear tetrahedron method was employed for all $k$ space integration with a $12 \times 12 \times 12$ subdivision in the full Brillouin zone for an ordered phase of $Ca_2MnOsO_6$. The magnetic ground state was obtained by computing the total energy of possible magnetic configurations. Our calculation was double-checked in selected cases using the full-potential linearized augmented plane wave method, as implemented in the WIEN2k code[29]. In order to estimate the correlation effects, the GGA+$U$ ($U$ is the Coulomb interaction) function with atomic-limit (AL) double-counting correction was used[30,31]. The values of $U$ were chosen to be 5 eV for Mn-3d and 1.5 eV for Os-5d, which were comparable to the values in the literature [9,32-36]. The self-consistent full-relativistic calculations were carried out with spin-orbit coupling (SOC) included. This is necessary to determine the magnetic anisotropy energy (MAE) and to check its influence on the electronic and magnetic properties. The calculation was first performed with the experimental lattice parameters obtained by SXRD, and then geometry optimization was conducted. The energy and charge convergence for the self-consistent calculation was set to $10^{-8}$ Hartree and $10^{-6}$ of electron, respectively.

## 3. RESULTS

## A. Crystal structure

The double perovskite oxide $A_2BB'O_6$ [$A$: alkali earth, $B$: 3d, $B'$: 4d(5d) elements] normally crystallizes into a monoclinic ($P2_1/n$), or tetragonal ($I4/m$), or cubic ($Fm$-$3m$) structure, in which two distinct octahedra $BO_6$ and $B'O_6$ align in a rock-salt fashion [37]. The degrees of the octahedral distortion and rotation seem to have a major impact on the lattice symmetry[38]. Based on the general trend, we attempted to characterize the crystal structure of $Ca_2MnOsO_6$ by applying these models. In particular, the monoclinic model ($P2_1/n$) was very effective to refine the SXRD pattern as reported for the analogous compounds such as $Ca_2FeOsO_6$ [7], $Ca_2CrOsO_6$ [2], and $Ca_2MnReO_6$ [39]. The refinement quality was, however, not fully satisfactory: a significant failure was found for the peaks marked by an arrow in the inset of Fig. 1. Although the major peaks were well refined by the $P2_1/n$ model (if Mn and Os are fully ordered), several expected peaks such as (0-11), (-101), and (101) were not observed above the background level. This failure suggested that the rock-salt-type ordered arrangement of $MnO_6$ and $OsO_6$ octahedra is not formed in $Ca_2MnOsO_6$[13], namely, Os and Mn atoms are distributed randomly over the perovskite $B$-site (i.e., anti-site disorder), as was found in $Ca_2MnRuO_6$ ($Pnma$ [17]).

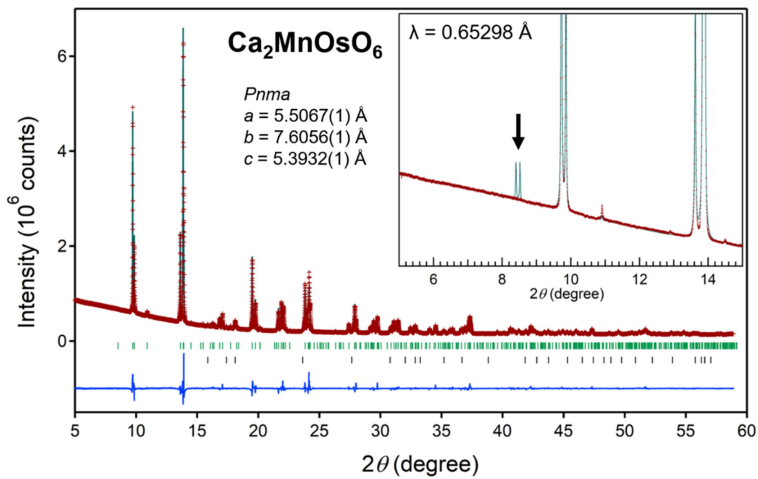

Fig. 1  Rietveld refinement of the powder synchrotron XRD pattern collected at room temperature. The crosses and solid lines show the observed and calculated patterns, respectively, with their difference shown at the bottom. The Os was analyzed simultaneously as the secondary phase. The estimated mass proportion was $Ca_2MnOsO_6$: Os = 0.986: 0.014. (Inset) Analysis by an ordered Mn

and Os atoms model; the arrow indicates an obvious discrepancy between the model and the experiment in the analysis.

We therefore tested the *Pnma* model of the SXRD data, and as shown in the main panel of Fig. 1, the *Pnma* model successfully characterized the structure of $Ca_2MnOsO_6$. The final solutions of the lattice parameters, atomic coordinates, and temperature factors are listed in Table 1. Note that the orthorhombic model (*Pnma*) was proposed for both end compounds, $CaOsO_3$[40] and $CaMnO_3$[41] as well, which indicates that $Ca_2MnOsO_6$ is not a double perovskite, but a solid solution of them.

**Table 1 Atomic coordinates and temperature factors ($B$) for $Ca_2MnOsO_6$ at room temperature obtained from synchrotron XRD**

| Atom | Site | Occp. | $x$ | $y$ | $z$ | $B$ (Å$^2$) |
| --- | --- | --- | --- | --- | --- | --- |
| Ca | 4$c$ | 1 | 0.9523(3) | 0.25 | 0.0102(9) | 1.12(3) |
| Os/Mn | 4$b$ | 0.5/0.5 | 0 | 0 | 0.5 | 0.22(1) |
| O1 | 4$c$ | 1 | 0.0340(11) | 0.25 | 0.5718(10) | 1.03(8) |
| O2 | 8$d$ | 1 | 0.2106(7) | 0.4578(5) | 0.1850(7) | 1.03 |

Space group: *Pnma*; lattice constants $a$ = 5.50669(4) Å, $b$ = 7.60567(6) Å, and $c$ = 5.39322(4) Å; $Z$ = 2; $d_{cal}$ = 6.1946 g/cm$^3$; and the final $R$ indices are $R_p$ = 2.029% and $R_{wp}$ = 3.597%.

The average of the $M$–O ($M$ = Mn or Os) bond lengths of $Ca_2MnOsO_6$ is compared with that of related compounds; it is shorter than that of $CaOs^{4+}O_3$[40] and $LaMn^{3+}O_3$[42] and longer than that of $NaOs^{5+}O_3$[43] and $CaMn^{4+}O_3$[41] (see Table 2). The comparison suggests that the valence state of $M$ is intermediate between the two groups. In addition, we examined the impact of the Jahn–Teller (JT) distortion on the average structure through comparison of the distortion factors (defined as the ratio of the longest to the shortest $M$–O length of an octahedron) of the compounds. It is 1.084 for Ca$_2$MnOsO$_6$, which is much smaller than 1.142 for LaMn$^{3+}$O$_3$ and much greater than 1.004 for CaMn$^{4+}$O$_3$, 1.030 for CaOs$^{4+}$O$_3$, and 1.004 for NaOs$^{5+}$O$_3$. Although the comparison indicates that the JT distortion could contribute to the average structure of Ca$_2$MnOsO$_6$, it is extremely difficult to uniquely identify the exact charge distribution on Mn and Os of Ca$_2$MnOsO$_6$ from only an analysis of the powder SXRD pattern, because of the anti-site disorder.

**Table 2 Comparison of structural parameters of Ca$_2$MnOsO$_6$ and related compounds**

|  | LaMn$^{3+}$O$_3$ [42] | CaMn$^{4+}$O$_3$ [41] | Ca$_2$MnOsO$_6$ [this work] | CaOs$^{4+}$O$_3$ [40] | NaOs$^{5+}$O$_3$ [43] |
|---|---|---|---|---|---|
| Space group | Pnma | Pnma | Pnma | Pnma | Pnma |
| $a$ (Å) | 5.5367(1) | 5.279(1) | 5.50666(4) | 5.57439(3) | 5.38420(1) |
| $b$ (Å) | 5.7473(1) | 7.448(1) | 7.60562(6) | 7.77067(4) | 7.58038(1) |
| $c$ (Å) | 7.6929(2) | 5.264(1) | 5.39317(4) | 5.44525(3) | 5.32817(1) |
| $V$ (Å$^3$) | 244.8 | 207.0 | 225.9 | 235.9 | 217.5 |
| $M$–O1 (Å) | 1.9680(3) × 2 | 1.895 × 2 | 1.945(1) × 2 | 2.003(1) × 2 | 1.946(1) × 2 |
| $M$–O2 (Å) | 1.907(1) × 2 | 1.900 × 2 | 1.917(4) × 2 | 1.978(4) × 2 | 1.939(3) × 2 |
| $M$–O2 (Å) | 2.178(1) × 2 | 1.903 × 2 | 2.077(4) × 2 | 2.037(4) × 2 | 1.940(3) × 2 |
| Distortion [a] | 1.142 | 1.004 | 1.084 | 1.030 | 1.004 |
| <$M$–O> (Å) | 2.018 | 1.899 | 1.980 | 2.006 | 1.942 |
| $M$–O1–$M$ (°) | 155.48(2) | 158.6 | 149.6(2) | 151.7(1) | 153.9(2) |
| $M$–O2–$M$ (°) | 155.11(5) | 157.2 | 155.7(1) | 152.0(2) | 155.2(2) |

[a] Defined as the ratio of the longest to the shortest $M$–O length of the $M$O$_6$ octahedron

**B. X-ray absorption spectroscopy**

To investigate the valence states of Mn and Os, we conducted the Mn-$L_{2,3}$ and Os-$L_3$ XAS on Ca$_2$MnOsO$_6$ and compared the spectra to those of MnO[44], LaMnO$_3$[44], SrMnO$_3$[45], and Sr$_2$FeOsO$_6$[46] which serve as Mn$^{2+}$, Mn$^{3+}$, and Mn$^{4+}$, and Os$^{5+}$ references, respectively (see Fig. 2). The spectral weight of the $L_3$ white line shifts to higher energies by 1 eV or more as the Mn valence state increases by one: from $Mn^{2+}$ (MnO) to $Mn^{3+}$ (LaMnO$_3$) and further to $Mn^{4+}$(SrMnO$_3$)[44,45]. The similar energy position and multiplet spectral features of the Mn-$L_{2,3}$ of Ca$_2$MnOsO$_6$ and LaMnO$_3$ demonstrate the trivalent state of Mn in Ca$_2$MnOsO$_6$. Analogously, the Os-$L_3$ spectrum of Ca$_2$MnOsO$_6$ locates at the same energy as that of Ca$_2$FeOsO$_6$, indicating the pentavalent state of the Os[7,47], thereby fulfilling the charge balance requirement for the $Mn^{3+}$/$Os^{5+}$ state. Consequently, the valence state of Ca$_2$MnOsO$_6$ ($Mn^{3+}$, $Os^{5+}$) is different from that of the analogous compound Ca$_2$MnRuO$_6$, in which the mixed valent states of ($Mn^{3+}$, $Ru^{5+}$) and ($Mn^{4+}$, $Ru^{4+}$) are degenerate in energy[17].

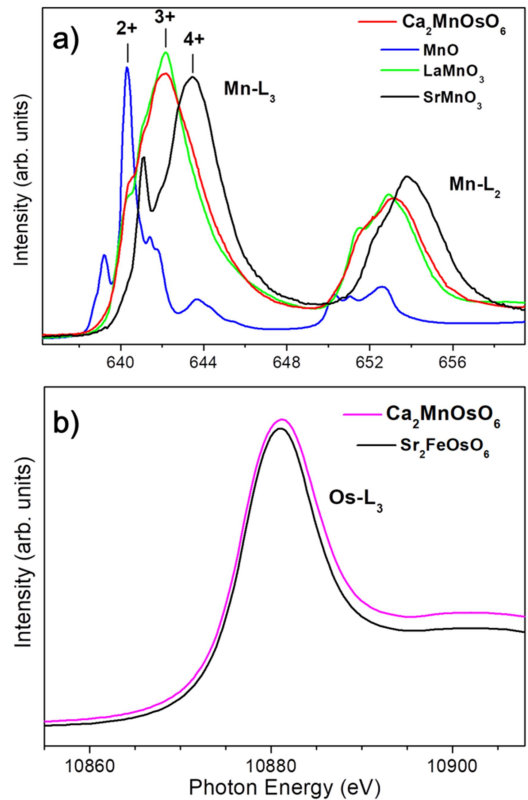

Fig. 2 (a) Room temperature Mn-$L_{2,3}$ XAS spectra of Ca$_2$MnOsO$_6$ and of MnO[44], LaMnO$_3$[44], SrMnO$_3$[45] as reference compounds for $Mn^{2+}$, $Mn^{3+}$, and $Mn^{4+}$ valence states. (b) Os-$L_3$ XAS spectra of Ca$_2$MnOsO$_6$ and of Sr$_2$FeOsO$_6$[46] as an $Os^{5+}$ reference.

**C. Electrical transport**

Fig. 3 shows an increasing resistivity $\rho(T)$ of $Ca_2MnOsO_6$ with lowering the temperature, experimentally demonstrating a semiconducting-like behavior. Interestingly, it is qualitatively and quantitatively different from the half-metallic behavior observed for the related compound $Ca_2MnRuO_6$[17,19]. The temperature dependence of $\rho(T)$ is poorly characterized by the Arrhenius model at low temperatures (< 300 K, Inset of Fig. 3); however, a variable-range-hopping conduction model better explains the data (inset of Fig. 3)[48]. Although the Arrhenius model does not exactly fit the data, we can roughly estimate the lower limit of the thermal activation energy in the high-temperature region (> 300 K) as ~0.11 eV from the linear fit as shown in the inset of Fig. 3.

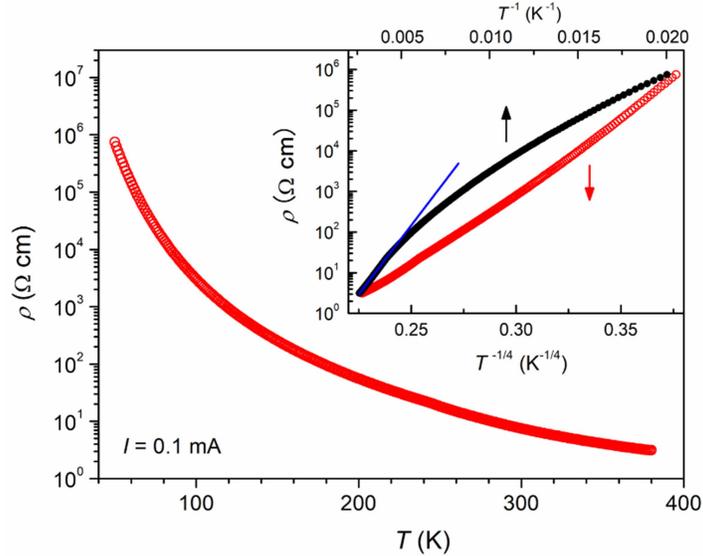

Fig. 3  Temperature dependence of $\rho$ for polycrystalline $Ca_2MnOsO_6$. The inset shows an alternative plot of the data. The blue line indicates a fitting to the Arrhenius law at high temperatures (> 300 K).

**D. Magnetic properties**

The magnetic susceptibility $\chi(T)$ curves measured in the zero field cooled (ZFC) and field cooled (FC) conditions for $Ca_2MnOsO_6$ are shown in Fig. 4. Upon cooling, there is an increasing value of $\chi$ at around 305 K, indicating the establishment of a FM-like transition. The divergence between the ZFC and FC curves at $T_c$ is negligible, but it becomes prominent at approximately 200 K, suggesting a possible formation of magnetic domains, or the like in the measurement process. We would like to note that the range of the high-temperature part ($> T_c$) of the $\chi(T)$ curve is too narrow for a meaningful Curie–Weiss analysis (see the right side of Fig. 4).

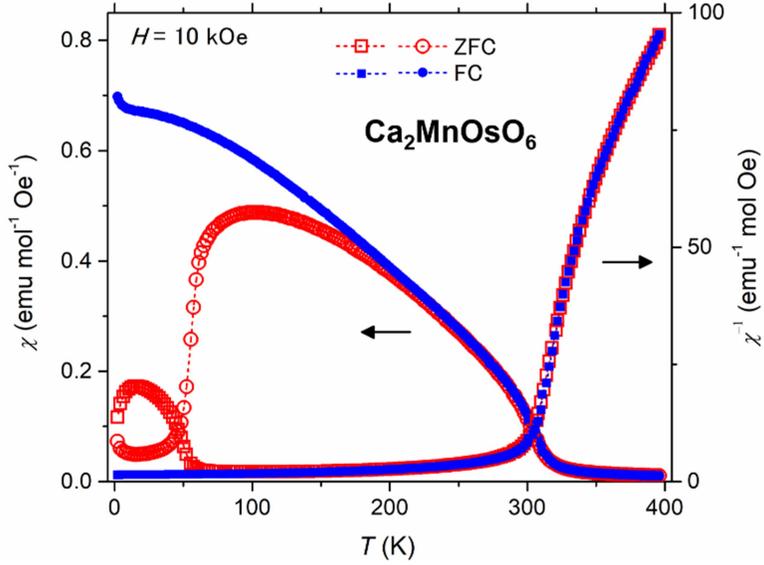

Fig. 4  Temperature dependence of $\chi$ of polycrystalline $Ca_2MnOsO_6$ measured in a field of 10 kOe. An alternative plot ($T$ vs. $\chi^{-1}$) is shown on the right side.

The AC $\chi$ ($= \chi' + i\chi''$) of $Ca_2MnOsO_6$ was measured at temperatures between 5 K and 320 K, and the $\chi'$ and $\chi''$ vs. $T$ curves are shown in Fig. 5. A sharp peak is observed at 305 K, which is consistent with the onset temperature of the $\chi(T)$ measurement. In contrast to multiple transitions found for $Ca_2MnRuO_6$, no additional anomaly is detected, indicating a single magnetic transition over the temperature range. Moreover, it substantiates that the divergence between the ZFC and FC curves in the $\chi(T)$ measurements is not caused by a magnetic transition. In Fig. 6, the temperature dependence of the heat capacity $C_p$ shows an anomaly evolving at 305 K in zero magnetic field, and it becomes broader when a magnetic field of 70 kOe is applied. This observation supports that this anomaly is caused by a magnetic transition.

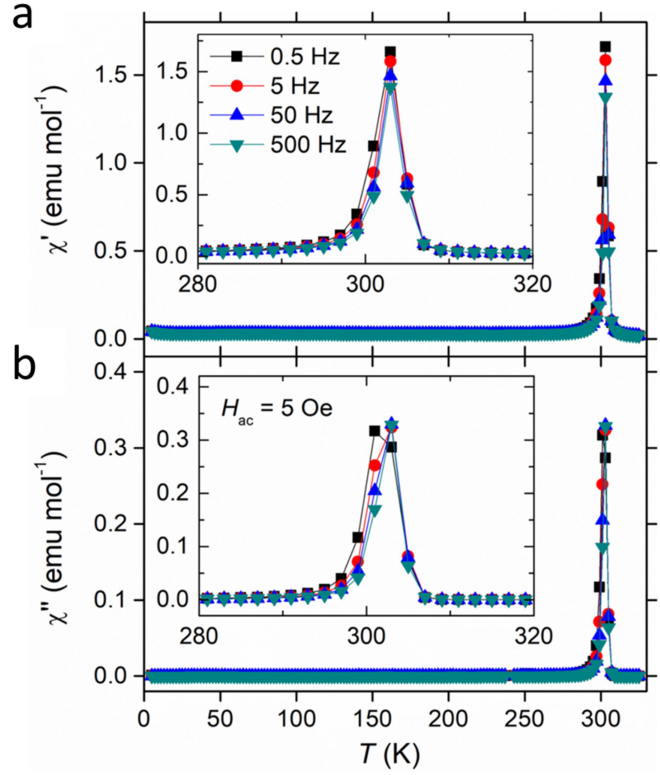

Fig. 5 (a) Real and (b) imaginary components of ac-magnetic susceptibility for polycrystalline $Ca_2MnOsO_6$ measured in an ac-magnetic field ($H_{ac}$) of 5 Oe at various frequencies. The inset shows a horizontal expansion around the magnetic transition temperature.

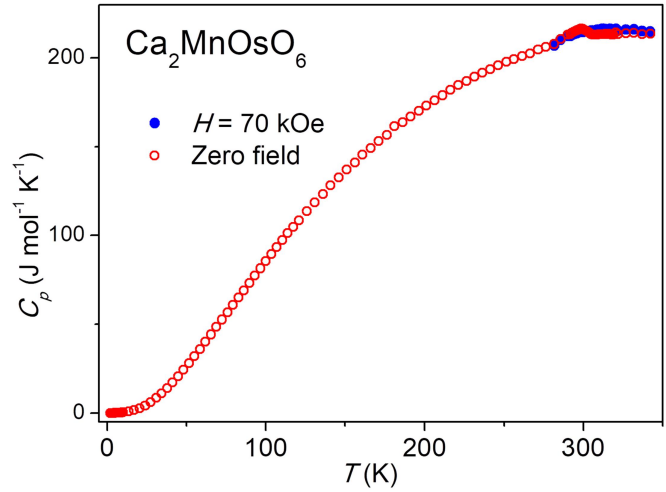

Fig. 6 Temperature dependence of specific heat capacity of $Ca_2MnOsO_6$.

The field-dependence of the magnetization was measured at several temperatures between 2

K and 350 K (see Fig. 7). The linear feature at 350 K confirms the paramagnetic behavior above $T_c$. In contrast, magnetic hysteresis appears at temperatures below $T_c$, suggesting that a FM component is involved in the magnetically ordered state. The magnetization at 2 K and 50 kOe is 1.40 $\mu_B$/f.u. which indicates the partial cancellation of the Mn and Os magnetic moments. The separate Mn and Os contributions to the total magnetic moment will be disentangled using the element specificity of X-ray magnetic circular dichroism (*vide infra*).

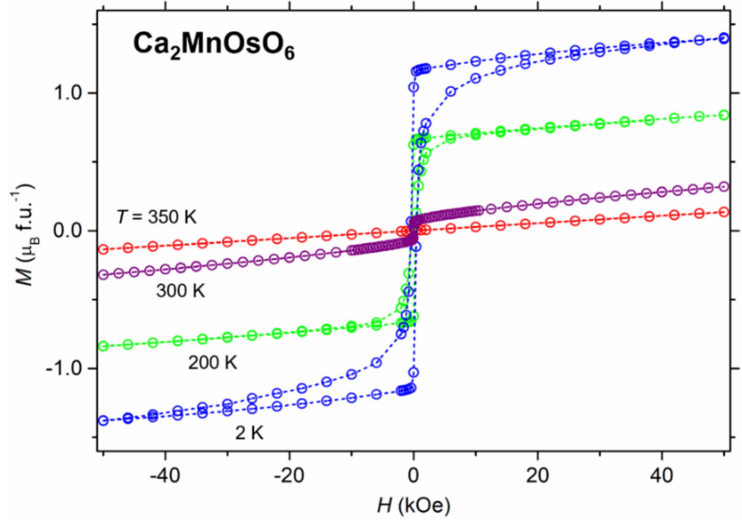

Fig. 7    Isothermal magnetization of polycrystalline $Ca_2MnOsO_6$ measured at various temperatures.

**E. X-ray magnetic circular dichroism**

Figure 8a shows the Mn-$L_{2,3}$ XAS spectrum (green) and the XMCD spectrum (blue) which is the difference between circularly polarized light with positive and negative helicities in an applied magnetic field of 60 kOe at 20 K. The XMCD signal can be clearly seen in Fig.8a. In order to extract the Mn moment, we performed the well-established configuration-interaction cluster calculations using the XTLS code[49]. The method uses a $MnO_6$ cluster, which includes explicitly the full atomic multiplet interaction, the hybridization of the Mn with the oxygen ligands, and the crystal field acting on the Mn ions. The hybridization strengths and the crystal field parameters were taken from Ref.[50].

Fig. 8b shows the calculated XAS (green) and XMCD (blue) for the Mn cluster in an exchange field ($H_{ex}$) of 30 meV. With this field (the energy scale of which reflects the experimentally determined $T_c$ of 305 K) the Mn is fully magnetized having a moment of 3.9 $\mu_B$. We observe that the line shape of the calculated XMCD spectrum is very similar to the experimental one, but we also notice that its magnitude is larger than measured. To reproduce the experimental size of XMCD signal in Fig.8a, we need to rescale the calculated XMCD spectrum by a factor of 1/3 as shown in Fig. 8b. This implies that the XMCD experiment presents the net Mn moment of 3.9/3 = 1.3 $\mu_B$.

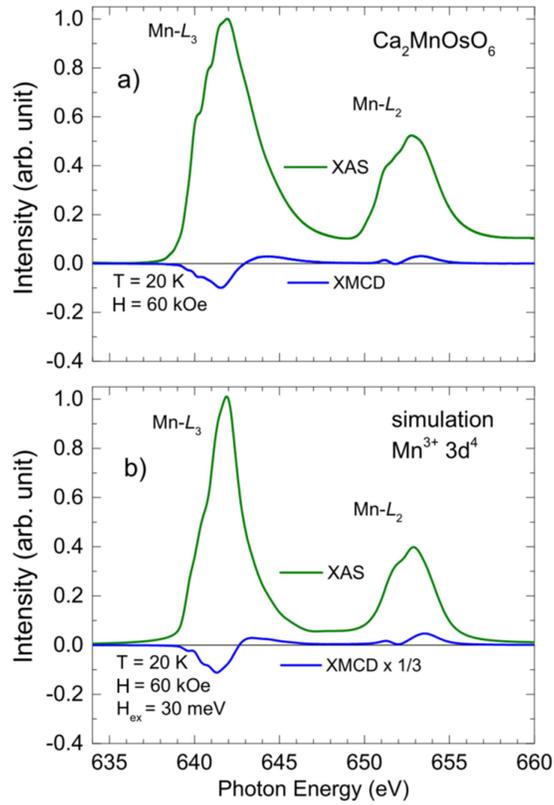

Fig. 8   (a) Mn-$L_{2,3}$ XAS (green) and XMCD (blue) spectra of Ca$_2$MnOsO$_6$ measured at 20 K under a magnetic field of 60 kOe and (b) the calculated XAS (green) and XMCD (blue) rescaled by a factor 1/3.

Figure 9 shows the Os-$L_{2,3}$ XAS spectrum (green curve) together with the XMCD spectrum (blue curve) measured below $T_c$ in an applied magnetic field of 13 kOe at 4 K. The XMCD signal of the Os has an opposite sign as that of the Mn, which indicates an antiparallel alignment of Os magnetic moment with that of the Mn. We have also performed configuration-interaction cluster calculations using an $OsO_6$ cluster with parameters taken from Ref.[51]. Fig. 9b shows the calculated XAS (green curve) and XMCD (blue curve) spectra. With an exchange field of 30 meV ($T_c$ of 305 K) the Os is fully magnetized having a moment of 2.05 $\mu_B$. The line shapes of the calculated spectra match very well those of the experiment. We notice, however, that the magnitude of the experimental XMCD is about 400 times smaller than that of the calculated, which suggests that most of the Os ions are magnetically disordered and/or antiferromagnetically aligned.

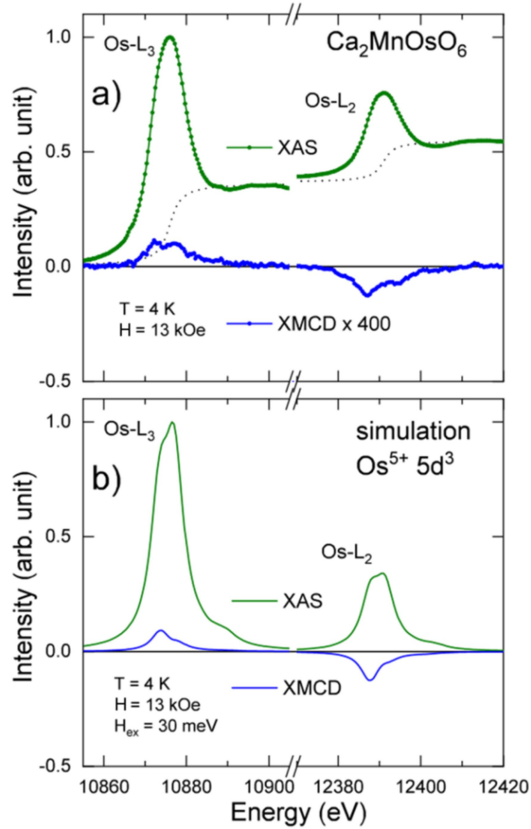

Fig. 9(a) Os-$L_{2,3}$ XAS (green) and XMCD (blue multiplied by 400) spectra of $Ca_2MnOsO_6$ measured at 4 K under a magnetic field of 13 kOe. (b) the calculated Os-$L_{2,3}$ XAS (green) and XMCD (blue) spectra.

We would like to note that our finding from XMCD that the net Mn moment is 1.3 $\mu_B$ and that the net Os moment is very small (antiparallel aligned) is in excellent agreement with the magnetization measurement (see the previous section) which yielded a total net moment of 1.40 $\mu_B$/f.u.

**F. Density functional theory (DFT) calculation**

We investigated the electronic and magnetic ground state of $Ca_2MnOsO_6$ by DFT calculations. In this theoretical study, we investigated a hypothetically ordered phase ($P2_1/n$; see Table S1, Figs. S1, and S2 in supplementary materials [52]). The symmetry of the $P2_1/n$ structure has been reduced further to lower symmetry P-1 (space group: 2) which gives rise to two in-equivalent atoms each of Os and Mn, respectively. The magnetic ground state of the ordered phase was studied by calculating the total energy for each of the five different magnetic states, namely FM (FM-↑↑↑↑; Os1: Os2: Mn1: Mn2), two AFM (AFM-1-↑↓↑↓ and AFM-2-↑↓↓↑), and two FIM (FIM-1-↑↑↓↓ and FIM-2-↑↑↑↓) states (see Fig. 10). From the total energy calculations, FIM-1 is found to have the lowest energy, with an energy difference of 198 meV/unit cell to that of the next lowest order (AFM-1). The spin-polarized DFT calculations for the hypothetically ordered $Ca_2MnOsO_6$ suggest that the magnetic ground state is certainly FIM-1 with the anti-parallel alignment of Mn moments to Os. The electronic ground state is HM (the HM gap is approximately 0.87 eV in the spin-up channel) within the GGA functional. Additionally, SOC effects has been considered where the spins are allowed to ordered along the [100] and [001] direction to check its influence on the electronic and related properties. Significant difference was not found for the net moments for Mn and Os (see Table 3), while a reasonable change has been observed in the electronic behavior.

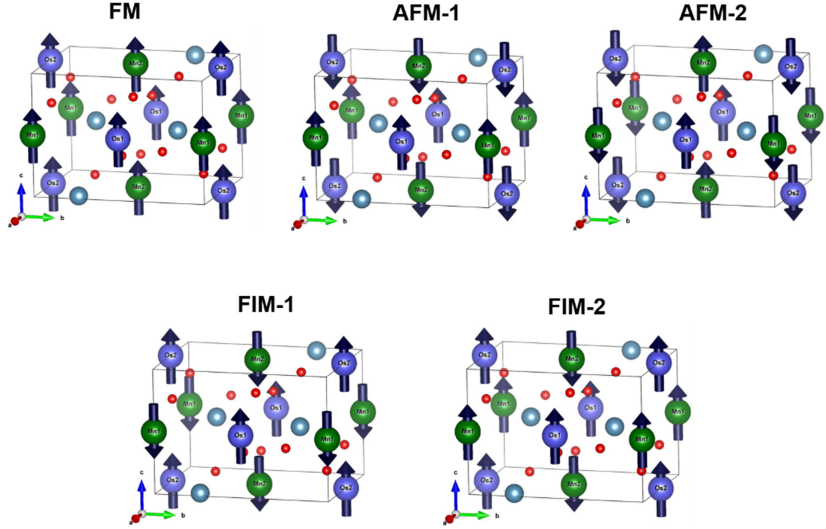

Fig. 10 Possible spin arrangements considered in the theoretical calculation for $Ca_2MnOsO_6$ with ordered Mn and Os atoms; ferromagnetic (FM), antiferromagnetic (AFM), and ferrimagnetic (FIM) configurations.

Table 3 Spin $<m_s>$ and orbital $<m_l>$ moments measured by XMCD and two sets of theoretical values for ordered $Ca_2MnOsO_6$

|  | XMCD [a] |  | GGA+SO [b] | | | |
|---|---|---|---|---|---|---|
|  | | | Ordered structure [100] | | Ordered structure [001] | |
|  | Mn | Os | Mn | Os | Mn | Os |
| $m_s$ ($\mu_B$/atom) | 1.3 | -0.005 | 3.13 | -1.38 | 3.14 | -1.41 |
| $m_l$ ($\mu_B$/atom) | 0.0 | 0.000 | 0.02 | 0.02 | 0.01 | 0.08 |
| $m_s+m_l$ ($\mu_B$/atom) | 1.3 | -0.005 | 3.15 | -1.36 | 3.16 | -1.33 |

[a] $T = 4$ K and $H = 13$ kOe for Os and 20 K and 60 kOe for Mn

[b] Standard generalized-gradient approximation with spin-orbit interaction

The electronic band structure and the spin-resolved DOS (total and partial) within GGA functional are shown in Fig. 11 for the ordered phase, which is HM with an insulating band gap of ~0.87 eV in the spin-up channel and metallic state in the spin-down channel. The major contributions to the total DOS around the $E_F$ are attributed to the Os-5d orbitals in both spin channels, which hybridize strongly with the O-2p orbitals. From the partial DOS and band structure (see Fig. 11 and Fig. S3a [52]), the Os-5d orbitals in the spin-up channel are found to be fully occupied by 5d-$t_{2g}$ orbitals, which shows the $Os^{5+}$ ($5d^3$) state. The six bands at and around the Fermi level ranging from -1.5 eV upto ~ +0.7 eV are from the Os-5d-$t_{2g}$ orbitals which hybridizes strongly with the O-2p orbitals (see Fig. 11 and Fig. S3a [52]) in spin-up and spin down channel. On the other hand, Mn-3d states are found to dominate mostly above +0.7 eV in the conduction region in spin-up channel, while in spin-down channel three d-orbitals (d-$t_{2g}$) are fully occupied lying below -1.5 eV while two of the d-states (d-$e_g$) cross the Fermi level signaling the partial occupancy in spin-down channel. This feature is close to the ionic picture $Mn^{3+}$ which should have four d-states occupied. With SOC taken into account (within GGA), band splitting was observed (see Fig. S3a [52]), however, the overall band gap did not open. This suggests the necessity of using GGA+U.

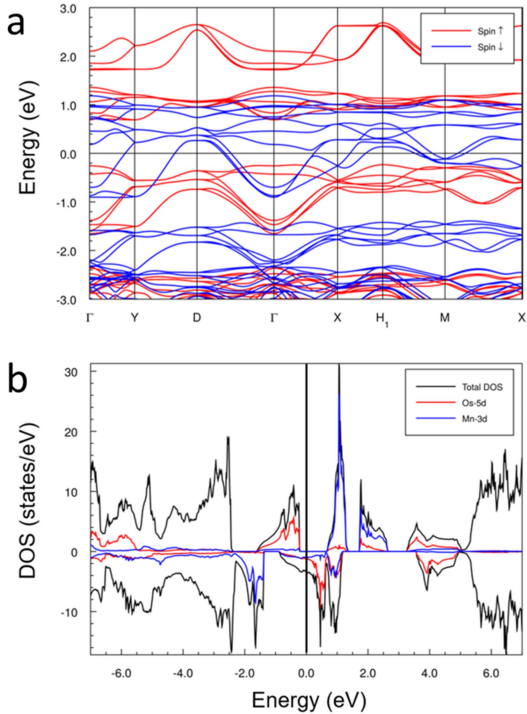

Fig. 11 (a) Band dispersion and (b) DOS of $Ca_2MnOsO_6$ with fully ordered Mn and Os atoms within GGA (optimized).

To further investigate the experimentally observed semiconducting transport, we performed a GGA+$U$ calculation with $U$ values ranging from 0–5 eV for Mn and 0–2.5 eV for Os[32-35]. The HM state changed to a semiconducting state at $U$(Mn) = 4 eV and $U$(Os) = 1.25 eV or higher[19]. With $U$ = 5 eV for Mn and 1.5 eV for Os, the calculated DOS and band structure are shown in Fig. 12. A band gap of 0.11 eV was achieved, which agrees well with the experiment (~0.11 eV). With SOC, an indirect band gap of 0.14 eV is found between the high symmetry point Γ and M, and a direct band gap of 0.21 eV at Γ point in the band dispersion shown in Fig. S3b [52]. Splitting of the orbitals for the easy axis [001] is large with a gap size of 0.14 eV. From the scalar relativistic DOS and band structure shown in Fig. 12, and relativistic fat bands in Fig. S3b [52], the Os-5d states are found to fully occupy with three $t_{2g}$ orbitals in the spin-up channel and the remaining $e_g$ orbitals are unoccupied, while in the spin-down channel, all the d-bands lie above $E_F$. Most of the bands from Mn-3d around $E_F$ hybridize with the O-2p and Os-5d states. The broad band lying just above $E_F$ is mainly contributed by the Mn-3d orbital; the exchange energy splitting is of the order of ~1.75 eV. Besides, MAE has been considered for the FIM-1 ground state. The calculated MAE is 29.6 meV/unit cell with its easy axis along the cubic [100] direction[53] within GGA. In contrast, within GGA+$U$, the easy axis is found along the out-of-plane with MAE of 3.46 meV/unit cell with all the spins aligned along the [001] direction.

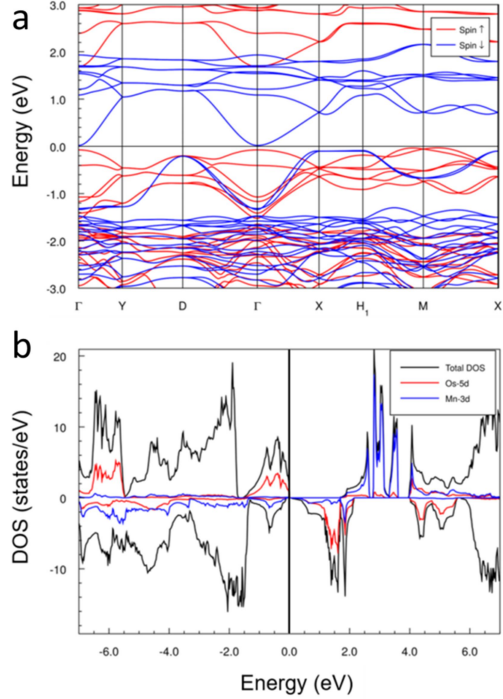

Fig. 12 (a) Band dispersion and (b) DOS of $Ca_2MnOsO_6$ with fully ordered Mn and Os atoms within GGA+$U$ ($U$ = 5 eV for Mn and 1.5 eV for Os atoms, respectively).

To evaluate the impact from anti-site disorder on the electronic ground state, five different anti-site-disordered configurations (AS1, AS2, AS3, AS4, and AS5) were created as shown in the Fig. S4 [52], where a 1 x 1 x 2 supercell has been generated which corresponds to a total of 40 atoms in a unit cell. DFT calculations were first carried out on these five different anti-site-disorder cases with FIM spin state. However, in comparison with the ordered FIM-1 case these anti-site-disordered cases showed higher energies within ~0.6 to 1.37 meV/unit-cell (see Table S2 [52]), with energies that AS1 < AS2 < AS3 < AS4 < AS5. We further calculated these anti-site-disordered cases with FM spin state. Energy differences (ΔE) between FM and FIM states in the respective anti-site-disordered cases are summarized in Table S3 [52]. FIM states showed lower energies, indicating that AFM alignment is favorable between Mn-Os, while FM coupling may be favorable among the Mn-Mn or Os-Os, giving rise to long-range FIM order in the anti-site-disordered $Ca_2MnOsO_6$. The electronic band gap is noted for AS1 and AS2 cases, while others remain metallic within GGA+U (see table S2 [52]).

**DISCUSSION**

We synthesized an anti-site-disordered double perovskite $Ca_2MnOsO_6$ (equivalent to $CaMn_{0.5}Os_{0.5}O_3$) with $Mn^{3+}$ and $Os^{5+}$ ions randomly distributed at the *B* site of the perovskite. It is noteworthy that neighbor compounds $Ca_2FeOsO_6$ and $Ca_2CrOsO_6$ all crystallize in ordered double perovskite structure with rocksalt arrangement of $FeO_6/CrO_6$ and $OsO_6$ octahedra. The *B* and *B'* site ordering in double perovskite is generally determined by the charge difference and effective ionic size difference of *B* and *B'* ions[54]. Given that $Mn^{3+}$ has the same effective ionic radii (0.645 Å) as that of $Fe^{3+}$ in octahedral coordination[55], the reason for the absence of $Mn^{3+}$ and $Os^{5+}$ ordering in $Ca_2MnOsO_6$ is unclear, although perhaps the JT distortions induced around the $Mn^{3+}$ ions may play an important role. Despite the absence of $Mn^{3+}$ and $Os^{5+}$ ordering, $Ca_2MnOsO_6$ is electrically semiconducting and features a remarkably high $T_c$ above room-temperature (305 K).

To rationalize the magnetic properties, we first consider the double perovskite $Ca_2MnOsO_6$ with perfect $Mn^{3+}$ and $Os^{5+}$ ordering. Our band structure calculations suggest that this would lead to a FIM ground state, where the moments of $Mn^{3+}$ ($t_{2g}^3 e_g^1$) and $Os^{5+}$ ($t_{2g}^3$) are arranged antiparallel. In order to interpret this result, we notice that the nearest-neighbor $Mn^{3+}$–O–$Os^{5+}$ super-exchange coupling can either be FM (the $Mn^{3+}$ $e_g$ electrons hop on to the empty $Os^{5+}$ $e_g$ states) or AFM (the down spin $Os^{5+}$ $t_{2g}$ electrons hop on to the $Mn^{3+}$ $t_{2g}$ states). In comparison to 3d ions, the 5d ions have much enhanced crystal field splitting, therefore the FM coupling involving the virtual hopping between $e_g$ orbitals becomes weak. In particular when the crystal structure is distorted this FM coupling can become even weaker[47,56], with the result that the AFM $Mn^{3+}$–O–$Os^{5+}$ coupling is the dominant interaction and generate the FIM state in the hypothetically ordered $Ca_2MnOsO_6$.

Introducing now anti-site disorder in $Ca_2MnOsO_6$, we will have also NN $Os^{5+}$–O–$Os^{5+}$ and $Mn^{3+}$–O–$Mn^{3+}$ interactions. The $Os^{5+}$–O–$Os^{5+}$ coupling is AFM as shown in AFM $NaOsO_3$[43,57]. The $Os^{5+}$ moments are therefore expected to compensate each other in such a situation. This would then be consistent with the experimental observation that the Os XMCD effect is very small. The case concerning the $Mn^{3+}$–O–$Mn^{3+}$ interactions is more complex. Depending on the orbital orientation due to the JT distortion of $Mn^{3+}$, the interaction can be either FM or AFM which has been well discussed in $LaMnO_3$[42,58-61]. In our crystal structure analysis, the contribution from the JT effect of $Mn^{3+}$ was clearly noticed (see crystal structure part). The moment, 1.3 $\mu_B/Mn^{3+}$, obtained from Mn XMCD is not negligible, but smaller than the theoretically calculated 3.15 $\mu_B/Mn^{3+}$ for ordered $Ca_2MnOsO_6$ phase (see Table 3) and also smaller than the 3.9 $\mu_B/Mn^{3+}$ from the cluster calculations. This indicates that there are competing FM and AFM $Mn^{3+}$–O–$Mn^{3+}$ interactions. Studies of highly anti-site disordered double perovskite $A_2Mn^{3+}B'O_6$ ($A$ = Ca, Sr; $B'$ = Ta, Sb), in which only $Mn^{3+}$ is magnetic, found that the FM interactions are dominant in these compounds with positive Weiss temperatures varied from 64–107 K [62]. A large FM component with a magnetization of about 1.47 $\mu_B/Mn^{3+}$ (at 5 K and 50 kOe) was observed in $Sr_2MnTaO_6$ despite its glassy state at low temperatures [62]. These results are thus similar to our $Ca_2MnOsO_6$ case.

Recent theoretical and experimental studies of the anti-site disorder on the magnetic properties of $Sr_2FeMoO_6$, $Sr_2FeReO_6$, $Sr_2CrReO_6$, $Sr_2CrOsO_6$, and $Sr_2FeRuO_6$ suggested that the anti-site disorder suppresses $T_c$ or destroys the magnetically ordered state, because the anti-site disorder introduces strong AFM $Fe^{3+}$–O–$Fe^{3+}$ and $Cr^{3+}$–O–$Cr^{3+}$ interactions to hinder the long-range magnetic order[11-14,63-65]. Different from $Fe^{3+}$ and $Cr^{3+}$, the $Mn^{3+}$–O–$Mn^{3+}$ super-exchange interactions can be both FM and AFM[42,58-61] depending on the orbital orientation, which may help to maintain net partial $Mn^{3+}$ moments and also a high $T_c$ in this anti-site-disordered $Ca_2MnOsO_6$. This scenario is likely to be different from the ferrimagnetism of anti-site-disordered $Ca_2MnRuO_6$[17,18], where the presence of mixed valence states of $Mn^{3+}/Mn^{4+}$–$Ru^{4+}/Ru^{5+}$ would also lead to a ferromagnetic double exchange mechanism for the $Mn^{3+}$–$Mn^{4+}$ and a HM state for $Ca_2MnRuO_6$[17-19].

## 4. CONCLUSION

Anti-site-disordered $Ca_2MnOsO_6$ was synthesized for the first time under high-pressure (6 GPa). It crystallizes into an orthorhombic structure (space group: *Pnma*), in which trivalent Mn and pentavalent Os share the Wycoff 4*b* position without an ordered arrangement. $Ca_2MnOsO_6$ is electrically semiconducting. XAS measurement confirmed the trivalent Mn and pentavalent Os oxidation states. The XMCD reveals the antiparallel alignment of the net Mn and Os magnetic moments. Remarkable is that the net Mn moment is only about 1/3 of its full $Mn^{3+}$ value and that the net Os moment is very small. We have discussed the strength and sign of various inter-site exchange interactions in this material using data from band structure calculations, taking into account also the presence of anti-site disorder and JT distortions around the $Mn^{3+}$ ions. The $T_c$ = 305 K is the second highest in the material category of so-called disordered ferromagnets and could therefore be useful in the development of oxide spintronic devices that are less sensitive for anti-site disorder during fabrication.

## ACKNOWLEDGMENTS

MPG thanks the Alexander von Humboldt Foundation for the financial support through HERMES program and Ulrike Nitzsche for the technical support. This study was supported in part by JSPS KAKENHI Grant Number JP16H04501, a research grant from Nippon Sheet Glass Foundation for Materials and Engineering (#40-37), and Innovative Science and Technology Initiative for Security, ATLA, Japan. We acknowledge support from the Max Planck-POSTECH-Hsinchu Center for Complex Phase Materials. The work in Dresden was partially supported by the Deutsche Forschungsgemeinschaft through SFB 1143 (project-id 247310070). The synchrotron radiation experiments were performed at the NIMS synchrotron X-ray station at SPring-8 with the approval of

the Japan Synchrotron Radiation Research Institute (Proposal Numbers 2017A4503, 2017B4502, 2018A4501, and 2018B4502). MG acknowledges support of NSF-DMR-1507252 grant of USA.